\documentstyle[aps,prl,preprint,floats,epsfig]{revtex}

\begin{document}

\preprint{\tighten\vbox{\hbox{\hfil CLNS 00/1696}
                        \hbox{\hfil CLEO 00-20}
}}

\title{Search for a scalar bottom quark with mass 3.5--4.5 GeV/$c^2$}  

\author{CLEO Collaboration}
\date{\today}

\maketitle
\tighten

\begin{abstract} 
We report on a search for a supersymmetric $\tilde{B}$ meson with mass
between 3.5 and 4.5 GeV/$c^2$ using 4.52 ${\rm fb}^{-1}$ of integrated
luminosity produced at $\sqrt{s}=10.52$ GeV,
just below the $e^+e^-\to B\bar{B}$ threshold, and collected with
the CLEO detector.  We find no evidence for a light scalar bottom quark.
\end{abstract}
\newpage

{
\renewcommand{\thefootnote}{\fnsymbol{footnote}}

\begin{center}
V.~Savinov,$^{1}$
T.~E.~Coan,$^{2}$ V.~Fadeyev,$^{2}$ Y.~S.~Gao,$^{2}$
Y.~Maravin,$^{2}$ I.~Narsky,$^{2}$ R.~Stroynowski,$^{2}$
J.~Ye,$^{2}$ T.~Wlodek,$^{2}$
M.~Artuso,$^{3}$
% ...
\end{center}
 
\small
\begin{center}
$^{1}${Stanford Linear Accelerator Center, Stanford University, Stanford,
California 94309}\\
$^{2}${Southern Methodist University, Dallas, Texas 75275}\\
$^{3}${Syracuse University, Syracuse, New York 13244}\\
% ...
\end{center}

\setcounter{footnote}{0}
}
\newpage

There has recently been renewed interest in the possibility of
a light scalar bottom quark~\cite{more-theory}, which ALEPH has
searched for~\cite{aleph}.
It has been noted that such a squark can exist in certain regions of
parameter space~\cite{carena}.
In addition, the production cross section
of scalar quarks in $e^+e^-$ annihilations, well above threshold,
is $\frac{1}{4}$ that of spin-$\frac{1}{2}$
quarks of the same charge; thus $\tilde b \bar{\tilde b}$ production would
contribute $\frac{1}{12}$ unit to $R$, the ratio of hadronic cross section
to $\mu^+\mu^-$ cross section,
and such an increase cannot be ruled out by existing measurements~\cite{R}.
Indirect evidence from the measured $B$ semileptonic branching fraction
disfavors the existence of a light $\tilde b$~\cite{nierste-plehn}.

If $\tilde b$ is the lightest supersymmetric particle (LSP),
and if $R$-parity is conserved, then $\tilde b$ would be stable.
If, instead, a scalar neutrino $\tilde\nu$ is the LSP, then $\tilde b$ will
decay $\tilde b \rightarrow c \ell^- \tilde\nu$ and/or
$\tilde b \rightarrow u \ell^-\tilde\nu$.
If $R$-parity is violated, $\tilde b$
will decay $\tilde b \rightarrow c \ell^-$ and/or
$\tilde b \rightarrow u \ell^-$. We have searched for a light $\tilde b$ that
decays $\tilde b \rightarrow c \ell^- \tilde \nu$ and/or
$\tilde b \rightarrow c \ell^-$.  Such a particle would dress itself as a
supersymmetric $\tilde B$ meson.  The dressed decays would be
$\tilde B \rightarrow D X \ell^- \tilde \nu$ and
$\tilde B \rightarrow D X \ell^-$, where $X$ represents possible additional
conventional hadrons.

The decays we search for, characterized by leptons and charmed
mesons, have much in common with conventional $B$ decays.
We perform a direct search, avoiding the $B$ background by using
a data sample collected below the $B \bar B$ threshold,
at $\sqrt{s} = 10.52$ GeV.
Our search covers the $\tilde B$ mass range 3.5--4.5 GeV/$c^2$.
Because our search
is near $\tilde B\bar{\tilde B}$ threshold, the production cross section cannot
be predicted with great precision.  We include the $\beta^3$ threshold factor
and ignore strong interaction effects to obtain
$\sigma(e^+e^-\to\tilde b\bar{\tilde b})=N_c\pi\alpha^2 Q_f^2\beta^3/3s$
(derived from Eq. 35.12 of Ref.~\cite{xsec}), where
$Q_f$, the charge of the $\tilde b$, is $-\frac{1}{3}$.

Below $B\bar{B}$ threshold, the major source of leptons and charmed mesons
is $e^+e^-\to c\bar{c}$,
which predominantly produces two charged leptons when both charmed mesons decay
semileptonically.  To suppress this $c\bar{c}$ background, we search for
events with two oppositely charged
leptons as well as a fully reconstructed hadronic $D$ or $D^*$ meson decay,
where $D^{(*)}$ denotes either a $D^{(*)0}$ or $D^{(*)+}$ meson.
Other sources of leptons include kaons that decay in flight,
photon conversions, and $\pi^0$ Dalitz decays.
A search for wrong-sign $D^{(*)}$-lepton
combinations was also conducted, but the resultant upper limits  
were significantly weaker than for the $D^{(*)}$-dilepton signature, and
they are not discussed further in this paper.

The data sample used in this analysis was produced in symmetric $e^+e^-$
collisions at the Cornell Electron Storage Ring (CESR)
and collected with the CLEO detector in two configurations, known as
CLEO II~\cite{cleonim} and CLEO II.V~\cite{iivnim}.
It comprises 4.52 ${\rm fb}^{-1}$ of integrated
luminosity produced at an 
$e^+e^-$ center-of-mass energy of $\sqrt{s}=10.52$ GeV,
35 MeV below the $B\bar{B}$ threshold.  Analysis of an additional
2.24 ${\rm fb}^{-1}$ collected on the $\Upsilon(4S)$ resonance provides
a sample of $B$ mesons, which can mimic the behavior of $\tilde{B}$ mesons,
thus allowing us to verify our experimental technique.  
The response of the experimental apparatus is studied with a
GEANT-based~\cite{geant} simulation of the CLEO detector, where the
simulated events are processed in a fashion similar to data.

In CLEO II,
the momenta of charged particles are measured with a
tracking system consisting of a six-layer straw
tube chamber, a ten-layer precision
drift chamber, and a 51-layer main drift chamber, all operating
inside a 1.5 T superconducting solenoid.  The main drift chamber
also provides a measurement of specific ionization energy loss,
which is used for particle identification.
For CLEO II.V, the six-layer straw tube chamber was replaced by a three-layer
double-sided silicon vertex detector, and the gas in the main 
drift chamber was changed from an argon-ethane to a helium-propane mixture.
Photons are detected with a 7800-crystal CsI electromagnetic calorimeter,
which is also inside the solenoid.
Proportional chambers placed at various depths within the steel return
yoke of the magnet identify muons.

Charged tracks are required to be well-measured and to satisfy track quality
requirements based on the average hit residual and the impact parameters
in both the $r$-$\phi$ and $r$-$z$ planes.  Muon candidates must
penetrate the steel absorber to a depth of at least five nuclear
interaction lengths, which effectively places a lower bound on the muon
momentum of 1.2 GeV/$c$.
Electrons are identified by a likelihood that includes the fraction of the
particle's energy deposited in the calorimeter and the spatial distribution of
the deposited energy.  To reduce contamination from low-momentum hadrons,
the electron
momentum is required to be greater than 1.0 GeV/$c$.  For electrons that are
combined with another lepton satisfying the above criteria,
the momentum requirement is lowered to 600 MeV/$c$.
Electron pairs from photon conversions are rejected by requiring the
dielectron invariant mass to be greater than 200 MeV/$c^2$.
We rely on the detector simulation to estimate the remaining contribution
of hadrons to the lepton sample.
$\pi^0$ candidates are formed from
pairs of photons with an invariant mass within 2.5 standard deviations of
the known $\pi^0$ mass that are kinematically fitted with the mass
constrained to the known $\pi^0$ mass.

$D$ mesons are reconstructed in the modes $D^0\to K^-\pi^+$ and
$D^+\to K^-\pi^+\pi^+$, where a sum over charge conjugate modes is implied.
The daughters of the $D$ candidates must undergo ionization energy loss
consistent with the particle hypothesis at the level of three standard
deviations.  To maximize detection efficiency, no requirements are placed
on the momentum of the $D$ candidate.  We reconstruct $D^*$ mesons in the
modes $D^{*0}\to D^0\pi^0$, $D^{*+}\to D^0\pi^+$, and $D^{*+}\to D^+\pi^0$.
When used in the $D^*$ modes, $D$ candidates must have invariant masses
less than 15 MeV/$c^2$ (about 2.5 standard deviations) from the known mass.

\begin{figure}[ht]
\begin{center}
\epsfig{file=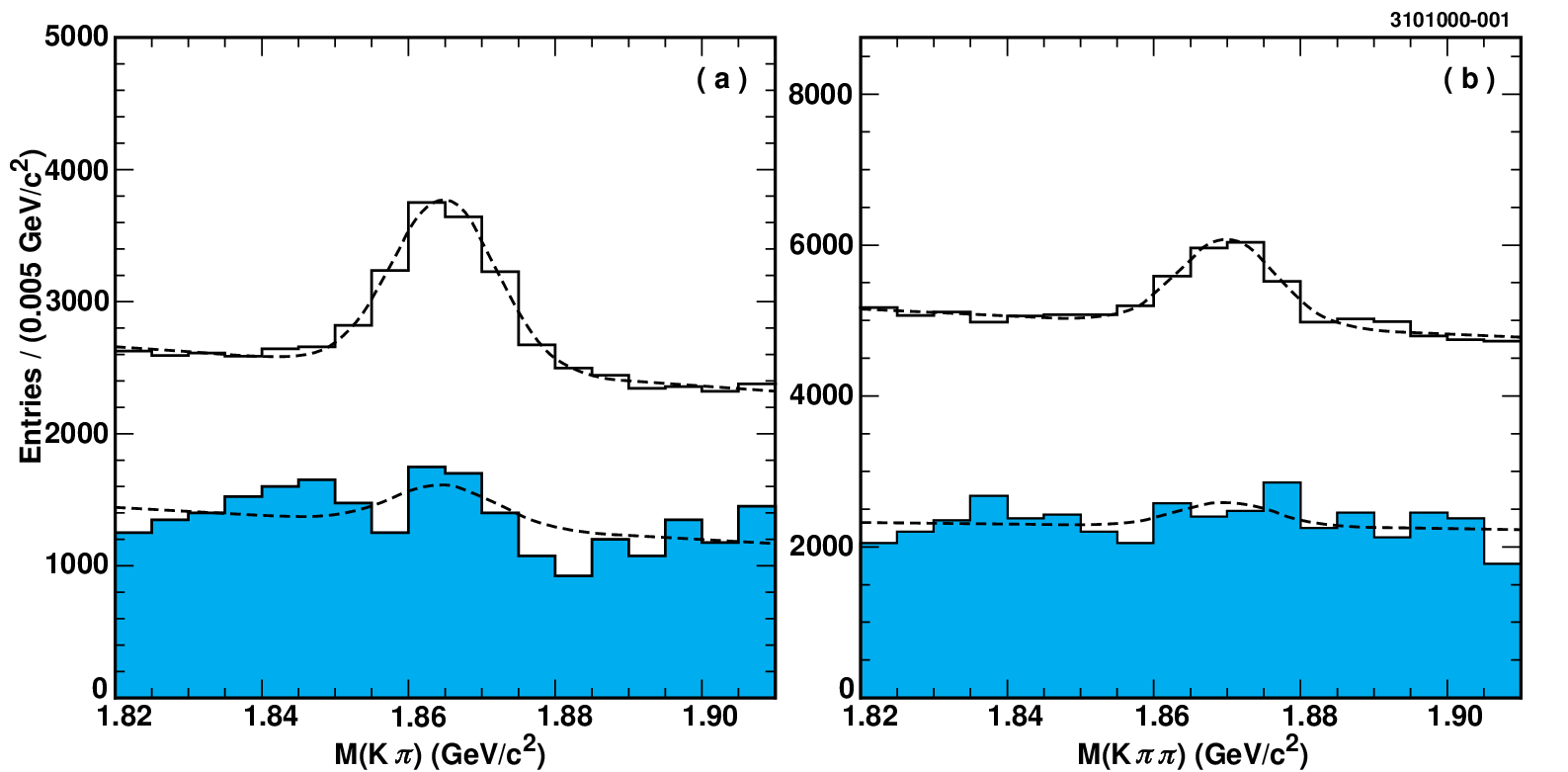, width=16.5cm}
\caption{Invariant mass distributions for
(a) $D^0\to K^-\pi^+$ and (b) $D^+\to K^-\pi^+\pi^+$
candidates paired with a single lepton of either charge (open histogram)
or with two oppositely charged leptons (shaded histogram).
The two $D$-dilepton distributions have been scaled by a factor of 25 to
facilitate comparison with the normalizing distributions.
The fits of each distribution to
a Gaussian and a linear background are shown in the dashed curves.}
\label{fig:dll}
\end{center}
\end{figure}

\begin{figure}[ht]
\begin{center}
\epsfig{file=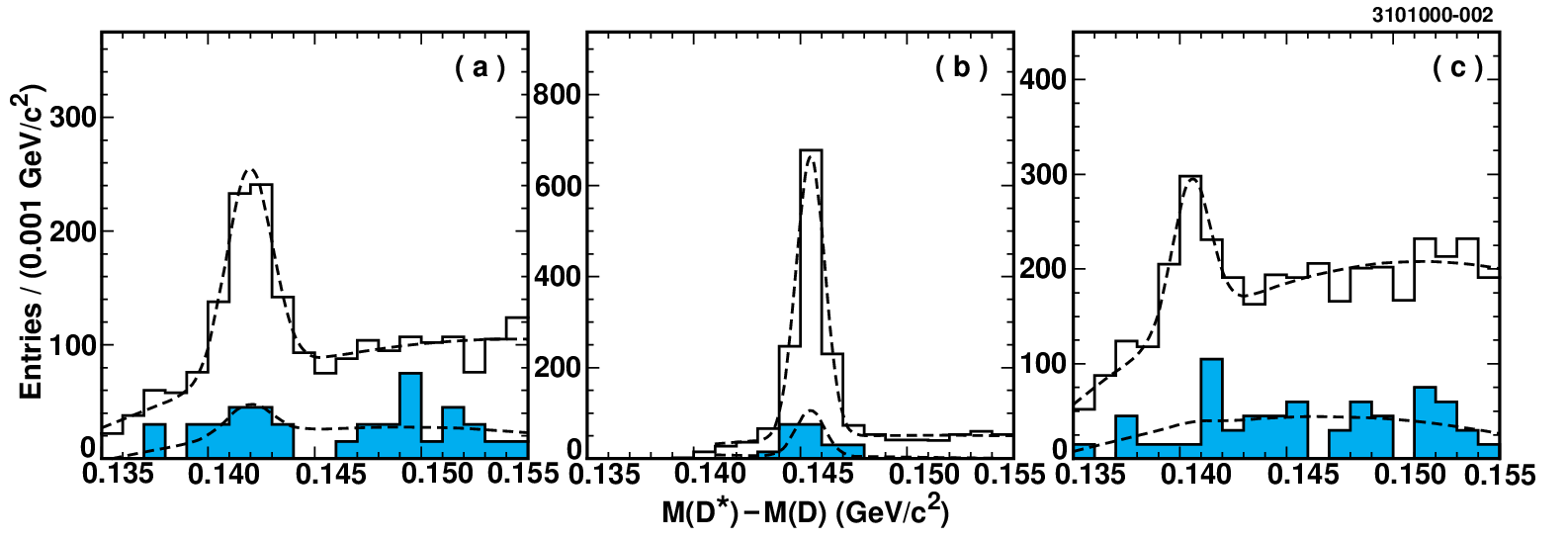, width=16.5cm}
\caption{Distributions of $M(D^*)-M(D)$ for
(a) $D^{*0}\to D^0\pi^0$,
(b) $D^{*+}\to D^0\pi^+$, and (c) $D^{*+}\to D^+\pi^0$
candidates paired with a single lepton
of either charge (open histogram) or with two
oppositely charged leptons (shaded histogram).
The three $D^*$-dilepton distributions have been scaled by a factor of 15 to
facilitate comparison with the normalizing distributions.
The fits of each distribution to
a Gaussian and a quadratic background are shown in the dashed curves.}
\label{fig:dstarll}
\end{center}
\end{figure}

Figure~\ref{fig:dll} shows the $K^-\pi^+$ and $K^-\pi^+\pi^+$
invariant mass distributions of $D^0$ and $D^+$ candidates, respectively,
associated with two oppositely charged leptons.
Also shown are the normalizing distributions for $D$ candidates
paired with a single lepton of either charge.
The analogous distributions for the $D^*$ modes, showing
the $D^*$-$D$ mass difference, are given in Figure~\ref{fig:dstarll}.
The $D^{(*)}$-dilepton distributions reveal a striking absence of signal.
The $D^{(*)}$ yields are extracted
from a fit of each histogram to a Gaussian distribution
over a linear background for the $D$ modes and a quadratic background for
the $D^*$ modes.
The means and widths of the Gaussians for the
$D^{(*)}$-dilepton fit are fixed to values determined from the normalizing
distributions, where the Gaussian widths are typically 6.5--7.0 MeV/$c^2$
for $M(D)$ and 0.5--1.0 MeV/$c^2$ for $M(D^*)-M(D)$.
Table~\ref{table:yields} lists the yields observed in data as well as
the expected
yields determined from a simulation of $e^+e^-\to q\bar{q}$ events, where
$q\in \{u,d,s,c\}$, hereafter referred to as ``continuum'' events.
No significant excess is observed in any mode.

In the normalizing distributions there is good
agreement between the data and simulated continuum events.
In addition, by analyzing data taken on the $\Upsilon(4S)$ resonance,
we observe the closely related semileptonic
$B$ decay with yields that
agree well with predictions from the detector simulation.

We determine the efficiency for detecting a supersymmetric $\tilde B$ meson
using Monte Carlo simulation.  A range of values for the $\tilde B$
mass, $M(\tilde B)$,
from 3.5 to 4.5 GeV/$c^2$, has been explored.  
Also, the sneutrino mass, $M(\tilde\nu)$, was varied within the kinematically
allowed range.
In the process
$e^+ e^- \rightarrow \tilde b \bar{\tilde b}$, followed by hadronization,
the supersymmetric $\tilde B$ mesons have energies $E(\tilde B)$ distributed
from $M(\tilde B)$ to $E_{\rm beam}$ according to some unknown
fragmentation function, which we approximate by a delta function
$\delta(E(\tilde B) - E_0)$, with $E_0$ given a value between $M(\tilde B)$
and $E_{\rm beam}$.  We simulate both $\tilde B$ mesons and their daughters,
as well as the additional hadrons that result from fragmentation.
We determine the dependence of efficiency on $M(\tilde B)$, $M(\tilde\nu)$,
and $E_0$.
To simulate the decay $\tilde B \rightarrow D \ell^- X$, we have taken $X$ to
be a single pion and used three-body phase space for the decay.  For
$\tilde b \rightarrow c \ell^- \tilde \nu$, we have used $D \ell^- \tilde \nu$
three-body phase space.

These variations affect the $\tilde{B}$ meson detection efficiency primarily
through the lepton momentum spectrum.
Table~\ref{table:yields} lists a representative set of detection efficiencies
and the resultant 95\% confidence level upper limits on the cross section,
assuming $M(\tilde{B})=4.0$ GeV/$c^2$, $M(\tilde\nu)=0$ GeV/$c^2$, 
and $E_0=E_{\rm beam}$.
The efficiency exhibits
weak dependence on $M(\tilde{B})$, but lower values of $E_0$
tend to soften the lepton momentum spectrum, resulting in reduced efficiency.
For the smallest value of $E_0$ we considered (3.7 GeV),
the efficiencies are one-fourth of those with $E_0=E_{\rm beam}$.
We give the upper limit on the product of the $\tilde{B}\bar{\tilde{B}}$
production cross section and the
$\tilde B\to D^{(*)}\ell\{\pi,\tilde\nu\}$ branching fraction
as a function of $M(\tilde{B})$ and $E_0$, assuming
$M(\tilde\nu)=0$ GeV/$c^2$.
For all modes, the variation of the efficiency with $M(\tilde\nu)$ is
roughly linear between $M(\tilde\nu)=0$ GeV/$c^2$ and
$M(\tilde\nu)_{\rm max} \equiv 0.6 M(\tilde B) - 0.8$ GeV/$c^2$,
where the efficiency vanishes.
Hence, the upper limits for a finite $M(\tilde\nu)$ would be
those for $M(\tilde\nu)=0$ GeV/$c^2$, scaled by
a factor $\left( 1-\frac{M(\tilde\nu)}{M(\tilde\nu)_{\rm max}}\right )^{-1}$.

The detection efficiency is also affected by the additional particles
produced in fragmentation.  From Monte Carlo studies, we have
determined that for each fragmentation particle, the fractional decrease
in efficiency is approximately 3\% and depends on the momentum and
species of the fragmentation particles.
By considering Monte Carlo simulation of $e^+e^-\to c\bar c$, we
estimate roughly two fragmentation particles per GeV of fragmentation energy.
The detection efficiency has been corrected accordingly.  A systematic
error per fragmentation particle of $\pm 2.5\%$ for the $D$ modes and
$\pm 5\%$ for the $D^*$ modes has been included in the upper limits to allow
for the uncertainty in this correction.

In addition, the upper limits have been inflated to account for uncertainties
in the detector simulation and in the $D$ and $D^*$ branching fractions,
which amount to systematic errors of 20\%.  We also assign an error of
10\% to the expected Standard Model yields due to undertainties
in the simulation of both fake and real leptons.  The total systematic
uncertainties are
small compared to the statistical errors on the background-subtracted yields.

\begin{figure}[ht]
\begin{center}
\epsfig{file=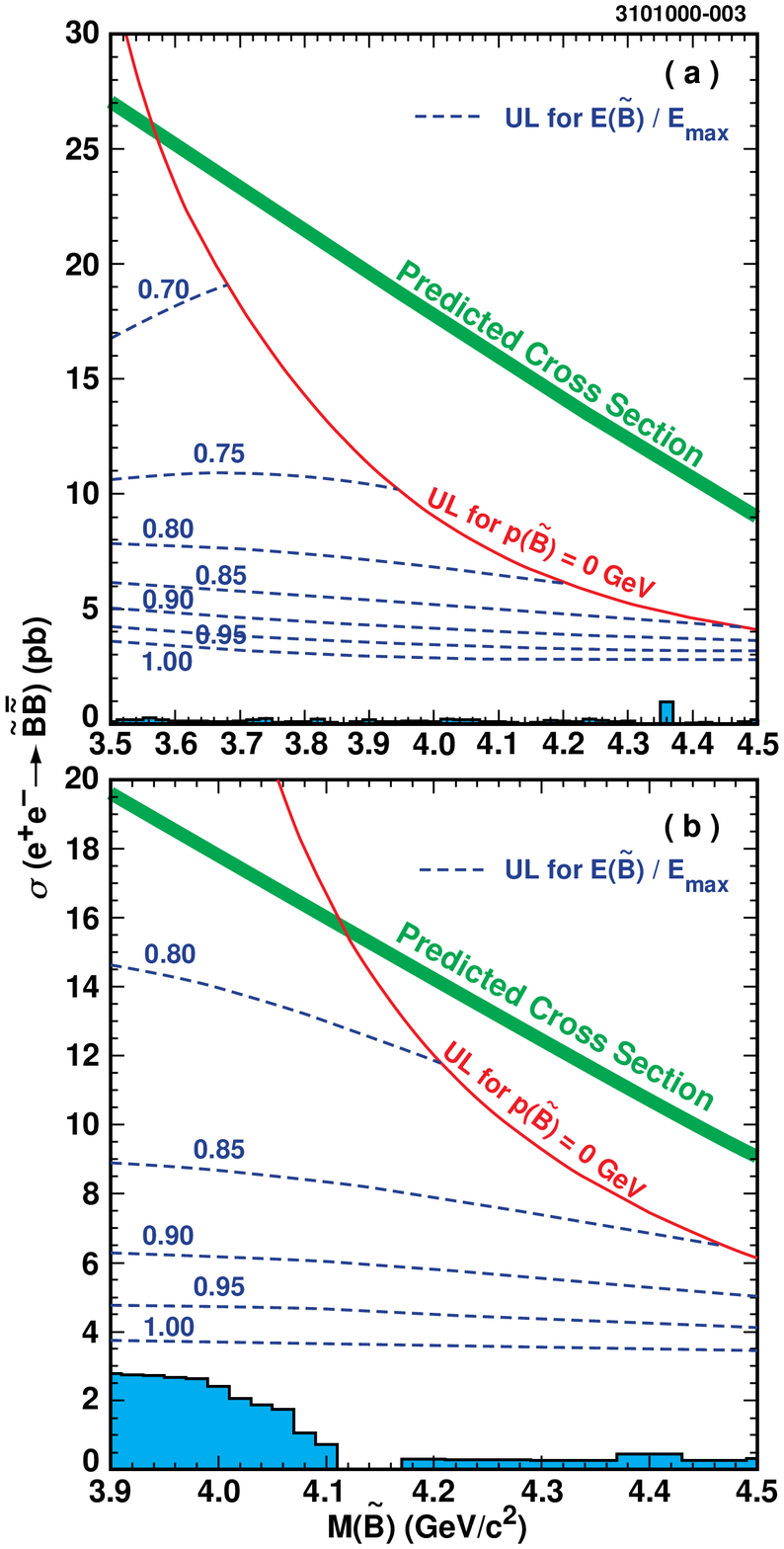,height=17cm}
\caption{
Upper limits at the 95\% confidence level on the product of the
$e^+e^-\to\tilde B\bar{\tilde{B}}$ production cross section
and
the $\tilde B\to D^{(*)}\ell^-\{\pi,\tilde\nu\}$ branching fraction.
The upper limits
for the (a) $D$-dilepton and (b) $D^*$-dilepton signatures
are given for the full kinematic range of $E(\tilde B)/E_{\rm beam}$
as a function of the $\tilde B$ mass, $M(\tilde B)$, assuming
$M(\tilde\nu)=0$ GeV/$c^2$.
The thick curves show the theoretically predicted cross
sections, derived from the cross section for pointlike fermions and scaled
by $\beta^3=\left ( 1-M(\tilde B)^2/E_{\rm beam}^2\right )^{3/2}$.
For finite $M(\tilde\nu)$, the upper limits are increased by a factor
$\left ( 1-\frac{M(\tilde\nu)}{0.6 M(\tilde B) - 0.8 \ {\rm GeV}/c^2}\right )^{-1}$.
The shaded histograms represent the largest upper limits, for each given
$M(\tilde B)$, on an $R$-parity-violating $\tilde B$ that decays
$\tilde B\to D^{(*)}\ell$.}
\label{fig:ul}
\end{center}
\end{figure}

Figure~\ref{fig:ul} shows the 95\% confidence level
upper limit on the product of
$\sigma(e^+e^-\to\tilde{B}\bar{\tilde{B}})$ and
${\cal B}(\tilde B\to D^{(*)}\ell\{\pi,\tilde\nu\})$
as a function of $M(\tilde{B})$ and $E_0$, assuming $M(\tilde\nu)=0$ GeV/$c^2$.
The strong rise of the upper limits with decreasing $E_0/E_{\rm beam}$
reflects the softening of the lepton spectrum.
Because of the $\beta^3$ dependence of the cross section near
threshold, the cross section at $\sqrt{s} = 10.52$ GeV is heavily dependent
on the assumed $M(\tilde{B})$. These predicted cross sections are shown as the
thick curves in Figure~\ref{fig:ul}.
For both the $D$-dilepton and $D^*$-dilepton signatures, the upper limits in
nearly all of the kinematically allowed $M(\tilde{B})$-$E_0$ parameter space
with $M(\tilde\nu)=0$ GeV/$c^2$
fall below the predicted $\tilde{b}\bar{\tilde{b}}$ production cross section.
Below $M(\tilde{B})=3.9$ GeV/$c^2$, the $D^*$-dilepton efficiency vanishes,
so we exclude a $\tilde B$ that decays $\tilde B\to D^*\ell\{\pi,\tilde\nu\}$
only in the mass range 3.9--4.5 GeV/$c^2$.
No portion of the parameter space is excluded if
$M(\tilde\nu)$ is greater than 1.2 GeV/$c^2$ and 1.3 GeV/$c^2$ for
the $D$-dilepton and $D^*$-dilepton signatures, respectively.

If the $R$-parity-violating decay $\tilde b\to c\ell^-$ were to occur,
then some fraction of the time it would appear as the dressed decay
$\tilde B\to D\ell^-$ or $\tilde B\to D^*\ell^-$.  We have examined the
$D\ell^-$ and $D^*\ell^-$ invariant mass distributions in
$D^{(*)}\ell^+\ell^-$ events for evidence of a peak indicating such
two-body decays.  We find no evidence of a peak.  Fitting the distributions to
a polynomial background plus a Gaussian with width given by our experimental
resolution (10 MeV/$c^2$), and stepping the Gaussian mean over the mass range,
we obtain upper limits on the product of the $\tilde b \bar{\tilde b}$
production cross section and the $\tilde B\to D^{(*)}\ell^-$ decay branching
fraction,
as a function of $M(\tilde B)$.  These upper limits, which include 15\%
systematic errors on the yields, are shown in Figure~\ref{fig:ul}.
For $\tilde B\to D\ell^-$, the upper limit is less than 0.3 pb for all masses
except near 4.36 GeV/$c^2$, where it is 1.0 pb.
For $\tilde B\to D^*\ell^-$ the upper limit is weaker, typically 3 pb
for masses around 4 GeV/$c^2$, dropping to 0.3 pb for masses near 4.5
GeV/$c^2$.

In conclusion, we have searched for associated $\tilde{B}\bar{\tilde{B}}$
production in $e^+e^-$ collisions with center-of-mass energy below the
$B\bar{B}$ production threshold.  We assume a branching fraction of 100\%
for $\tilde{B}\to D^{(*)}\ell^-\pi$ or  $\tilde{B}\to D^{(*)}\ell^-\tilde\nu$.
Considering $D$-dilepton and $D^*$-dilepton combinations,
we find no evidence of a light scalar bottom quark produced at
$\sqrt{s} = 10.52$ GeV.
Upper limits on the $\tilde{B}\bar{\tilde{B}}$ production cross section
depend on the assumed mass and energy of the $\tilde{B}$ meson, as well as
the mass of the sneutrino.
For $M(\tilde\nu)$ less than ${\cal O}(1 \ {\rm GeV}/c^2)$,
the existence of a light scalar bottom quark with mass between
3.5 GeV/$c^2$ and 4.5 GeV/$c^2$ has been excluded at the 95\% confidence level.
A light scalar quark decaying 100\% of the time 
$\tilde b\to c\ell\tilde\nu$ and/or $\tilde b\to c\ell$
would have escaped our notice only if its
decay matrix element results in a lepton spectrum
much softer than three-body phase space.

\begin{acknowledgements}
We thank Tung-Mow Yan and Matthias Neubert for useful discussions.
We gratefully acknowledge the effort of the CESR staff in providing us with
excellent luminosity and running conditions.
I.P.J. Shipsey thanks the NYI program of the NSF, 
M. Selen thanks the PFF program of the NSF, 
A.H. Mahmood thanks the Texas Advanced Research Program,
M. Selen and H. Yamamoto thank the OJI program of DOE, 
M. Selen thanks the Research Corporation, 
F. Blanc thanks the Swiss National Science Foundation, 
and H. Schwarthoff and E. von Toerne
thank the Alexander von Humboldt Stiftung for support.  
This work was supported by the National Science Foundation, the
U.S. Department of Energy, and the Natural Sciences and Engineering Research 
Council of Canada.
\end{acknowledgements}

\renewcommand{\baselinestretch}{1}

\begin{table}[ht]
\begin{center}
\caption{Fitted $D$ and $D^*$ yields
for the $D^{(*)}$-dilepton signature.  The expected yields due to Standard
Model processes are obtained from an
analysis of simulated continuum events and include a systematic error of
10\% due to uncertainties in the modeling of both fake and real leptons.
The representative signal efficiencies given include the $D^{(*)}$ branching
fractions and assume $M(\tilde{B}) = 4.0$ GeV/$c^2$, $E_0=E_{\rm beam}$, and
$M(\tilde\nu)=0$ GeV/$c^2$.
The corresponding 95\% confidence level upper limits on the product of the
$e^+e^-\to\tilde b\bar{\tilde b}$ production cross section and the
$\tilde B\to D^{(*)}\ell\{\pi,\tilde\nu\}$ branching fraction
are calculated from the excess in the measured yield over the expected yield.
These upper limits include systematic errors of 20\% on the reconstruction
efficiency.}
\begin{tabular}{ccccc}
Mode  &  Yield  &  Expected Yield &  $\epsilon_{sig}$ (\%) &
	$\sigma(e^+e^-\to\tilde{b}\bar{\tilde{b}})\times
	{\cal B}(\tilde B\to D^{(*)}\ell\{\pi,\tilde\nu\})$ \\
\hline
$D^0\ell^+\ell^-$ & $47.6\pm 20.0$ & $33.7\pm 14.7\pm 3.4$ & $0.44\pm 0.02$ \\
$D^\pm\ell^+\ell^-$ & $37.4\pm 25.1$ & $58.3\pm 21.3\pm 5.8$ & $0.86\pm 0.04$\\
$D\ell^+\ell^-$ & &&& $< 2.7$ pb at 95\% C.L.\\
\hline
$D^{*0}\ell^+\ell^-$ & ${\phantom 1}4.9\pm 3.4$ & $4.2\pm 2.6\pm 0.4$ & $0.08\pm 0.01$ \\
$D^{*\pm}(D^0\pi^\pm)\ell^+\ell^-$ & $11.3\pm 3.8$ & $4.0\pm 2.0\pm 0.4$ &
	$0.12\pm 0.01$ \\
$D^{*\pm}(D^\pm\pi^0)\ell^+\ell^-$ & ${\phantom 1}0.2\pm 3.6$ & $3.6\pm 2.3\pm 0.4$ &
	$0.09\pm 0.01$ \\
$D^*\ell^+\ell^-$ & &&& $< 3.7$ pb at 95\% C.L. \\
\end{tabular}
\label{table:yields}
\end{center}
\end{table}

\end{document}